\documentclass[intlimits,twoside,a4paper]{article}

\usepackage{graphicx}
\usepackage[cp1251]{inputenc}

\usepackage[eqsecnum]{cmpj3}

\issue{2017}{20}{2}{23703}
\doinumber{10.5488/CMP.20.23703}
\title[Photoionization cross section in a spherical quantum dot]%
{Photoionization cross section in a spherical quantum dot: Effects of some parabolic confining electric potentials%
}
\author[M. Tshipa]{M. Tshipa}
\address{University of Botswana, Corner of Notwane and Mobuto Road, P/Bag 00704, Gaborone,
Botswana }

\date{Received January 6, 2017, in final form April 13, 2017}
\begin{document}

\maketitle

\begin{abstract}

A theoretical investigation of the effects of spatial variation of confining electric potential on photoionization cross section (PCS) in a spherical quantum dot is presented. The potential profiles considered here are the shifted parabolic potential and the inverse lateral shifted parabolic potential compared with the well-studied parabolic potential. The primary findings are that parabolic potential and the inverse lateral shifted parabolic potential blue shift the peaks of the PCS while the shifted parabolic potential causes a red shift.

\pacs 71.55.Eq,	73.21.La,  73.22.Dj
\end{abstract}

\section{Introduction}
\label{intro}

Recent advances in nanofabrication technology have made it possible to fabricate nanostructures of different sizes and shapes \cite{Byer15,Zul15,Noro12,Yao09,Sere15,Zhao15}. Nanostructures have a wide range of applications including in nanomedicine \cite{Chen11}, information processing \cite{Haus10}, energy physics \cite{Pana15} and gas sensing \cite{Yang15}, to mention a few. When fabricating these structures, it is impossible to eliminate all impurities. In some cases, it may be advantageous to introduce impurities to improve the performance of nanodevices (doping). If the impurity happens to be positively charged, then an electron may be bound to it. Given enough energy, the electron can break free from the electrostatic grasp of the impurity. The excitation energy can be in different forms, one of which is electromagnetic radiation of appropriate frequency. In this regard, photoionization studies on nanostructures could offer an insight into the electron-impurity interaction in a wide variety of conditions. As such, the literature is awash with investigations of PCS \cite{Hash15,Xie14,Yil10,Necu11,Sahin08}.

The purpose of this work is to explore the effects of some of the potentials that vary parabolically with the radial distance on the PCS in a spherical quantum dot (SQD). This communication is organized as follows. The theory is presented in section~\ref{prenote}. The following section~\ref{res} entails discussions and analysis, and concluding remarks are laid in section~\ref{conc}.

\section{Preliminary notes}
	\label{prenote}

The system investigated is a hydrogenic impurity located at the centre of a spherical quantum dot of radius $R$, which may be a GaAs material embedded in a GaAlAs matrix. The electric potentials inside the spherical dot assume parabolic spatial variations with the radial distance from the centre of the quantum dot. Photoionization is a process in which a bound charge carrier is liberated to the continuum by some appropriate radiation, with cross section \cite{Xie14}
\begin{equation}
	\label{eq:pi}
	\sigma_{lm}=\sigma_{0}\hbar\omega\sum_{f}|\langle f|\vec{r}|i \rangle |^2\delta(E_f-E_i-\hbar\omega),
\end{equation}
where $\hbar\omega$  is the photon energy and $\sigma_{0}=(4\piup^2\alpha_{\text{FS}}n_rE_{\text{in}}^2)/(3\epsilon E_{\text{av}}^2)$. $E_{\text{in}}$ is the effective incident electric field, $E_{\text{av}}$ is the average electric field in the dot of refractive index $n_r$ and dielectric constant $\epsilon$. $E_i$ and $E_f$ are the energies associated with the initial and the final eigenstates $|i\rangle$ and $|f\rangle$, respectively. $\langle f|\vec{r}|i\rangle$ is the usual matrix element coupling the initial states to the final states, $\alpha_{\text{FS}}$ is the fine structure and $\vec{r}$ is the position vector. The energy conserving $\delta$ function is replaced by the Lorentzian function
\begin{equation}
	\label{eq:dl}
	\delta(E_f-E_i-\hbar\omega)=\frac{\hbar\Gamma}{\piup\{[\hbar\omega-(E_f-E_i)]^2+(\hbar\Gamma)^2\}}\,,
\end{equation}
where $\Gamma$ is the hydrogenic impurity linewidth.

Due to the spherical symmetry of the system, electron wave functions can be cast in the form $\Psi(\theta,\phi,r)=C_{lm}Y_{lm}(\theta,\phi)\chi(r)$. $Y_{lm}(\theta,\phi)$ are the usual spherical harmonics whereas $\chi(r)$ is the radial component of the wave function satisfying the Schr\"odinger equation, within the effective mass approximation,
\begin{equation}
	\label{eq:eq1}
	\frac{1}{r^2}\frac{\rd}{\rd r}\left[r^2\frac{\rd}{\rd r}\chi(r)\right]+\left\{\frac{2\mu}{\hbar^2}\left[E_{lm}+\frac{k_ee^2}{\epsilon r}-V(r)\right]-\frac{l(l+1)}{r^2}\right\} \chi(r)=0,
\end{equation}
where $\mu$ is the effective mass of the electron (of charge $-e$), $k_e$ is the Coulomb constant. Angular momentum and magnetic quantum numbers are designated by $l$ and $m$, respectively, and $C_{lm}$ is the normalization constant.

The confining electric potentials considered here are the parabolic, shifted parabolic and the inverse lateral shifted parabolic potentials, each superimposed on an infinite spherical quantum well (ISQW). For these calculations, the initial states are described by wave functions of bound electrons (with impurity) while the final states are associated with wave functions for an electron in an SQD without the charged impurity. Evaluation of the matrix elements for an SQD leads to the  selection rules $\Delta l = \pm1$. Therefore, the $l$ values of the final and initial states will differ by unity.

\subsection{Parabolic potential}
When the parabolic potential, which has the form
\begin{equation}
	\label{eq:ppot}
		V(r)=\frac{1}{2}\mu\omega_0^2r^2, \qquad  (r<R),
\end{equation}
is inserted into the Schr\"odinger equation~(\ref{eq:eq1}) in the presence of the donor impurity, then the second order differential equation is solvable in terms of the Heun Biconfluent function \cite{Ron95,Hort13}
\begin{equation}
	\label{eq:heun}
	\chi(r)=C_{1lm}\re^{g_1(r)}r^{l}\text{HeunB}\big(2l+1,\alpha,\beta,\gamma,g_2(r)\big)+C_{2lm}\re^{g_1(r)}r^{-(l+1)}\text{HeunB}
\big(-(2l+1),\alpha,\beta,\gamma,g_2(r)\big)
\end{equation}
with
 \begin{equation}
  \label{eq:abc}
  	\alpha=0,\qquad
	\beta=-\frac{2E_{lm}}{\hbar\omega_0}\,,\qquad
	\gamma=\frac{4k_ee^2}{\hbar\epsilon}\sqrt{-\frac{\mu}{\hbar\omega_0}}\,,
 \end{equation}
and the arguments
\begin{equation}
    \label{eq:g1}	
g_1(r)=\frac{\mu\omega_0}{2\hbar}r^2,\qquad
	g_2(r)=\sqrt{2g_1(r)}.
\end{equation}

For a solid quantum dot, the coefficient corresponding to the second linearly independent solution, $C_{2lm}$, should be taken as zero due to the divergent nature of this solution at the centre of the SQD. Applying the boundary condition, concerning the continuity of the wave function at the walls of the SQD ($r=R$), leads to the energy spectrum of an electron in an SQD with a parabolic potential as
\begin{equation}
  E_{lm}=-\frac{1}{2}\beta_R\hbar\omega_0\,,
\end{equation}	
with $\beta_R$ satisfying the condition
\begin{equation}
	\label{eq:bc}
		\text{HeunB}\big(2l+1,\alpha,\beta_R,\gamma,g_1(R)\big)=0.
\end{equation}
In the absence of the impurity, the radial part of the wave function is in terms of the hypergeometric function
\begin{equation}
	\label{eq:hyp}
	\chi^{0}(r)=C_{lm}^0\re^{-g_1(r)}r^{\frac{l}{2}-\frac{3}{4}}M\big([a^0],[b],2g_1(r)\big),
\end{equation}
where $C_{lm}^0=C_{1lm}^0$ is the normalization constant and
\begin{equation}
 	a^0=\frac{l}{2}+\frac{3}{4}-\frac{E_{lm}^0}{2\hbar\omega_0}\,, \qquad
	b=l+\frac{3}{2}\,.
\end{equation}

Here, $g_1(r)$ is given by equation~(\ref{eq:g1}). The demand that the wave function vanishes at the walls of the nanostructure yields the energy dispersion relation
\begin{equation}
	E_{lm}^{0}=\left(l+\frac{3}{2}-2a^0_R\right)\hbar\omega_0\,,
\end{equation}
where $a^0_R$ gives the condition $M\big([a^0_R],[b],2g_1(R)\big)=0$.

\subsection{Shifted parabolic potential}
This potential is convex: maximum at the centre, and decreases parabolically to assume a minimum value (here taken as zero) at the radius;
\begin{equation}
	\label{eq:sppot}
		V(r)=\frac{1}{2}\mu\omega_0^2(r-R)^2, \qquad  (r<R).
\end{equation}
The solution to the radial component of the Schr\"odinger equation~(\ref{eq:eq1}) corresponding to this potential is also in terms of the Heun Biconfleunt function [equation~(\ref{eq:heun})] but with
 \begin{equation}
 	\alpha=2\sqrt{-\frac{\mu\omega_0R^2}{\hbar}}\,,\qquad
	g_1(r)=\frac{\mu\omega_0}{2\hbar}(r-2R)r,\qquad
	g_2(r)=-\ri\sqrt{\frac{\mu\omega_0}{\hbar}}r.	
\end{equation}
$\beta$ and $\gamma$ are the same as those for the parabolic potential with the donor impurity. The energy spectrum is given by the usual boundary conditions at the walls of the SQD as
\begin{equation}
	E_{ml}=-\frac{\beta_R}{2}\hbar\omega_0\,,
\end{equation}
where $\beta_R$ is the value of $\beta$ that satisfies the condition given in equation~(\ref{eq:bc}).

Without the donor impurity, the solution to the radial part of the Schr\"odinger equation is of the same functional form as that in equation~(\ref{eq:heun}) with all the parameters having the same expressions as for the case with the impurity except for $\beta=\beta^0$ and $\gamma$ which are
$$\beta^0=-\frac{2E_{lm}^0}{\hbar\omega_0}$$ and $$\gamma=0.$$

The energy eigenvalues for a shifted parabolic confining potential without the impurity can be written as
\begin{equation}
	E_{lm}^0=-\frac{\beta_R^0}{2}\hbar\omega_0\,,
\end{equation}
where $\beta_R^0$ is the value of $\beta=\beta^0$ that satisfies equation~(\ref{eq:bc}).

\subsection{Inverse lateral shifted parabolic potential}

This potential has a concave increase in the radial distance from the centre of the SQD, with a maximum value at the walls of the SQD:
\begin{equation}
	\label{eq:ilsppot}
		V(r)=\frac{1}{2}\mu\omega_0^2[(2R-r)r], \qquad  (r<R).
\end{equation}

The radial wave functions corresponding to this potential configuration are given by equation~(\ref{eq:heun}), with \[\beta=\frac{(2E_{lm}-\mu\omega_0^2R^2)}{\ri\hbar\omega_0}\] and \[g_2(r)=\ri\left(-\frac{\mu^2\omega_0^2}{\hbar^2}\right)^{1/4}r,\]
with $\alpha$, $\gamma$ and $g_1(r)$ being identical to those for the shifted parabolic potential in the presence of the impurity. Again, imposing the boundary conditions at the walls of the SQD yields the equation for the determination of the energy eigenvalues;
\begin{equation}
	E_{lm}=\frac{\ri}{2}\beta_R\hbar\omega_0+\frac{1}{2}\mu\omega_0^2R^2,
\end{equation}
where $\beta_R$ is the value of $\beta$ that satisfies the condition set in equation~(\ref{eq:bc}).

In the absence of the impurity, the solution to the radial part of the Schr\"odinger equation is of the same form as in equation~(\ref{eq:heun}), but with $\gamma=0$ and
\[\beta=\beta^0=\frac{(2E_{lm}^0-\mu\omega_0^2R^2)}{\ri\hbar\omega_0}\,.\]

All other parameters remain the same as for the case of this potential with the donor impurity. The application of the boundary conditions gives the  energy eigenvalues in the absence of the impurity for an inverse lateral shifted parabolic potential as
\begin{equation}
	E_{lm}=\frac{\ri}{2}\beta_R^0\hbar\omega_0+\frac{1}{2}\mu\omega_0^2R^2,
\end{equation}
where $\beta_R^0$ is the value of $\beta^0$ that satisfies the condition set in equation~(\ref{eq:bc}).

\section{Main results}
\label{res}

\begin{figure}[!t]
\centerline{\includegraphics[width=0.54\textwidth]{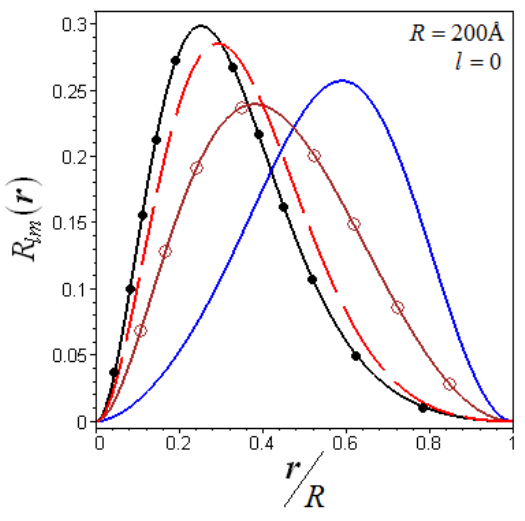}}
\caption{(Color online) The effect of different potentials on the normalised ground state radial probability density for an SQD of radius $R=200$~{\AA}.  The curve with circles represents an infinite spherical square well while the dashed corresponds to the parabolic potential, the solid plot is associated with the shifted parabolic potential and the inverse lateral shifted parabolic potential is represented by the solid line with dots, each of strength $\hbar\omega_0=20$~meV.}
\label{fig:rpd}
\end{figure}

The parameters used in these calculations are relevant to GaAs quantum dots: effective electronic mass $\mu=0.067m_e$, $m_e$ being the free electron mass and $\epsilon=12.5$. The impurity linewidth has been taken such that $\hbar\Gamma=0.2$~meV. Figure~\ref{fig:rpd} displays the effects of these potential geometries on the normalised ground state radial probability density [$R_{lm}(r)=r^2\chi(r)^2/R^2$] in the presence of the hydrogenic impurity for an SQD of radius $R=200$~{\AA}. The plots with circles represent an infinite spherical square well, the dashed curve (solid plot with dots) corresponds to the parabolic potential (inverse lateral shifted parabolic potential) while the solid plot represents the shifted parabolic potential. The parabolic potential and the inverse lateral shifted parabolic potential have the propensity to shift peaks of the radial probability density towards the centre of the SQD, which decreases the electron-impurity distance of separation. The influence of the inverse lateral shifted parabolic potential is slightly more pronounced than that of the parabolic potential because, even though the two are equal at the centre and at the walls of the SQDs, the former is always greater in the region between the centre and the walls. The shifted parabolic potential, on the other hand, shifts the peaks of the radial probability density towards the outer regions of the SQD, increasing the electron-impurity distance of separation.

\begin{figure}[!t]
\centerline{\includegraphics[width=0.54\textwidth]{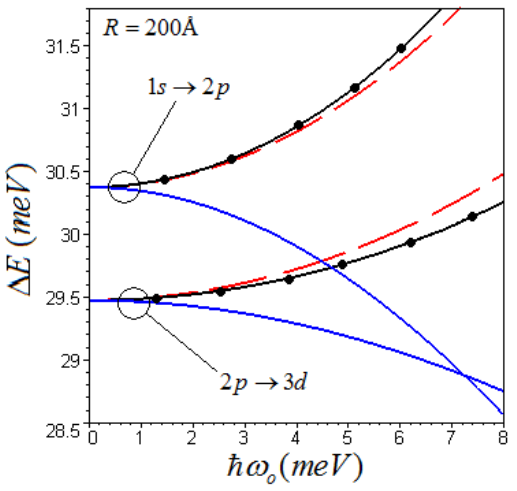}}
\caption{(Color online) The dependence of the first and second order transition energies on the strengths of different potentials. The solid (dashed) curves represent the inverse lateral shifted parabolic potential (parabolic potential), while the plots with dots are associated with the shifted parabolic potential, for $R=200$~{\AA}.}
	\label{fig:deo}
\end{figure}

The $l>0$ electrons are localized towards the walls of the SQD. As such, the radial position expectation values of such electrons are in the regions where both  parabolic and inverse lateral shifted parabolic potentials are higher than for the ground state. This gives these potentials the proclivity to affect the higher $l$ valued electrons more than the lower $l$ valued electrons. As a result, transition energies increase with increasing strengths of the potentials. On the other hand, the shifted parabolic potential decreases the transition energies. This is because this potential affects the lower $l$ valued electrons more than it does the higher $l$ valued electrons, since the lower $l$ valued electrons spend most of their time in the regions where this potential is greater. This is depicted in figure~\ref{fig:deo}, which shows transition energies as  functions of strengths of the three potentials, in an SQD of radius $200$~{\AA}. In the figure, the parabolic potential is represented by dashed plots, the solid curves are associated with the shifted parabolic potential while the inverse lateral shifted parabolic potential is represented by the solid plots with dots. Inevitably, the energy needed for photoionization is modified as the strengths of these potentials are varied. Thus, peaks of the photoionization cross section undergo shifts as strengths of these potentials increase, according to the dependence of the transition energies on the strengths of different potentials.

In the absence of the impurity, the first order transition energies ($s\rightarrow p$), $\Delta E_{sp}$, are less than those of the second order ($p\rightarrow d$), $\Delta E_{pd}$, for an ISQW for all values of nanodot radius. In the presence of the impurity, there exists a radius at which the first order and the second order transition energies coincide.  In these calculations, this radius is in the neighbourhood of $R_{0}=171$~{\AA}. For SQDs with radii less (greater) than $R_0$, the second order transition energies are more (less) than the first order transition energies. The parabolic and the inverse lateral shifted parabolic potentials reduce the value of this radius as they intensify. On the contrary, increasing the strength of the shifted parabolic potential increases $R_{0}$, sending it to infinity as it intensifies further. In this case, $\Delta E_{sp}$ and $\Delta E_{pd}$ would never coincide and $\Delta E_{pd} > \Delta E_{sp}$.

\begin{figure}[!t]
\centerline{\includegraphics[width=0.54\textwidth]{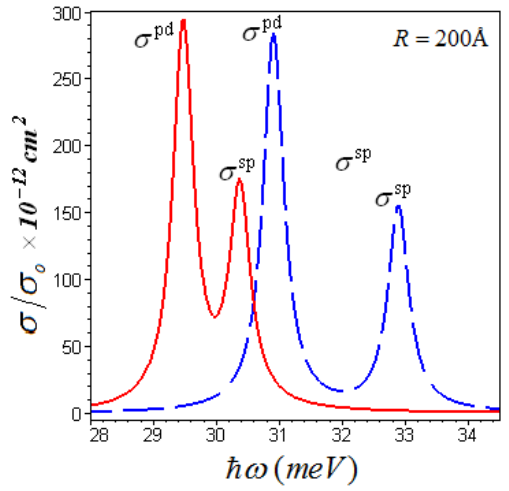}}
\caption{(Color online) The sum of the first and second PCSs as functions of beam energy for the ISQW
 ($\hbar\omega_0 = 0$~meV) (solid curve) and for the parabolic potential of strength $\hbar\omega_0 = 10$~meV (dashed curve) superimposed upon the ISQW,
 for an SQD of radius $R = 200$~\AA.}
\label{fig:cswop}
\end{figure}

\begin{figure}[!t]
\centerline{\includegraphics[width=0.54\textwidth]{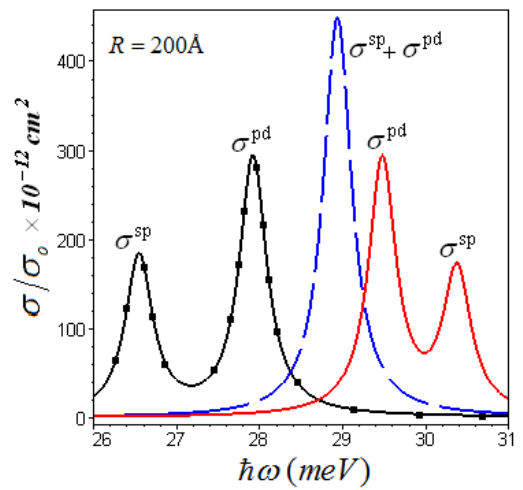}}
\caption{(Color online) The influence of the shifted parabolic potential on the variation of normalized PCS with the perturbing field energy for an SQD of radius $R=200$~{\AA}. The curves are for different potential strengths: $\hbar\omega_0=0$~meV (solid plot), $\hbar\omega_0=7$~meV (dashed curve) and $\hbar\omega_0=12$~meV (solid plot with squares), superimposed upon the ISQW.}
\label{fig:cswos}
\end{figure}
\begin{figure}[!b]
\centerline{\includegraphics[width=0.54\textwidth]{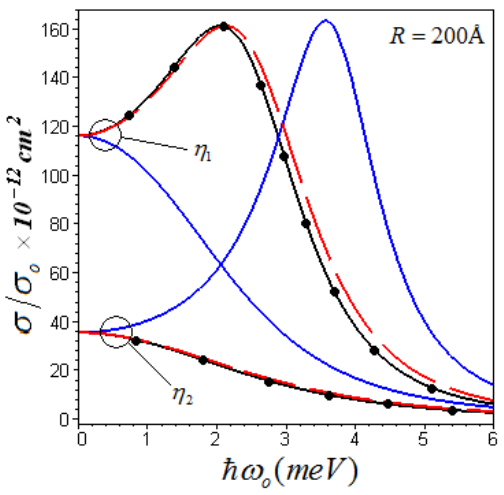}}
\caption{(Color online) The dependence of the $s\rightarrow p$ transition PCS on strengths of different potentials: the parabolic (inverse lateral shifted parabolic potential) is represented by the dashed curves (plots with dots) while the shifted parabolic potential is associated with the solid lines. The two bundles of graphs are generated for the beam energies $\eta_1=\hbar\omega=30$~meV and $\eta_2=\hbar\omega=30.5$~meV, all for an SQD of radius $200$~{\AA}.}
\label{fig:csow2}
\end{figure}

In figure~\ref{fig:cswop}, which depicts the $s\rightarrow p$ and $p\rightarrow d$ transitions PCSs, the SQD radius is greater than $R_0$, thus the $s\rightarrow p$ peak occurs at larger beam energies than the second order peak. Increasing strengths of the parabolic and the inverse lateral shifted parabolic potentials blue shifts the peaks of the PCS, simultaneously moving them apart (figure~\ref{fig:cswop}). This can be beneficial in cases where transitions between different states (for example, the $s\rightarrow p$ and the $p\rightarrow d$ transitions) need to be isolated and distinct, for research or practical purposes. The inverse lateral shifted parabolic potential is generally greater than the parabolic potential. Therefore, it has more propensity to blue shift and separate the PCS peaks than the latter. The shifted parabolic potential red shifts the peaks of the PCSs, with the first order peak experiencing more shifting until it equals and surpasses the $p\rightarrow d$ peak in being red shifted (figure~\ref{fig:cswos}). This implies that the shifted parabolic potential can be utilized to have the $s\rightarrow p$  and the $p\rightarrow d$ transitions having the same photon energies of excitation, or even to have control over which transition exactly is to have a lower photon energy of excitation. The PCS in figures~\ref{fig:cswop} and~\ref{fig:cswos} for an infinite cylindrical quantum well ($\hbar\omega_0=0$~meV) are concurrent with those in the literature \cite{Necu11}.

Figure~\ref{fig:csow2} shows the normalized PCSs for the $s \rightarrow p$ transition as functions of the strengths of the three confining electric potentials. The dashed plots correspond to the parabolic potential, the shifted parabolic potential is represented by the solid curves while the plots with dots are associated with the inverse lateral shifted parabolic potential. The PCSs are plotted for two beam energies, one is less than the ISQW transition energy $\Delta E_0$ (transition energies corresponding to $\hbar \omega_0 = 0$), while the other is slightly greater than $\Delta E_0$. Since the parabolic potential and the inverse lateral shifted parabolic potential enhance transition energies as they intensify (figure~\ref{fig:deo}), an increase in the strengths of the two potentials capacitates photoionization only when the beam energy is greater than $\Delta E_0$. Contrarily, an increase in the strength of the shifted parabolic potential will capacitate photoionization for beam energies that are less than $\Delta E_0$, owing to its proclivity to decrease transition energies (also in figure~\ref{fig:deo}).

\section{Conclusions}
\label{conc}

Electron states in a spherical quantum dot have been obtained within the effective mass regime, which were utilized to calculate the photoionization cross section associated with a centred donor impurity. In particular, the effect of intrinsic confining electric potentials that have parabolic dependence on the radial distance was interrogated. The primary findings are that the parabolic potential and the inverse lateral shifted parabolic potential blue shift the peaks of the PCS, while the shifted parabolic potential red shifts the peaks. Thus, the three electric potentials can be used to tune nanodevices, without necessarily having to alter the dimensions of the SQDs.

\subsection*{Acknowledgements}

The author wishes to humbly thank M. Masale and T.G. Motsumi for the invaluable discussions we have had.

\ukrainianpart

\title{Переріз фотоіонізації в сферичній квановій точці:  Ефекти деяких параболічних обмежуючих електричних потенціалів%
}
\author{M. Тшіпа}
\address{Університет Ботсвани, м. Габороне, Ботсвана}

\makeukrtitle

\begin{abstract}
Представлено теоретичне дослідження впливу просторової зміни обмежуючих електричних потенціалів на переріз фотоіонізації в сферичній квантовій точці. Розглянуті тут профілі потенціалів є зміщений параболічний потенціал і дзеркальне зображення зміщеного параболічного потенціалу, які порівнюються з добре вивченим параболічним потенціалом.  Основний результат --- це те, що параболічний потенціал і дзеркальне зображення зміщеного параболічного потенціалу зсувають піки перерізу фотоіонізації у фіолетовий діапазон, тоді як зміщений параболічний потенціал спричинює   червоний зсув.

\keywords переріз фотоіонізації, електричний обмежуючий потенціал, сферична квантова точка, воднеподібна домішка
\end{abstract}
  \end{document}